\documentclass{iau} 
\usepackage{graphicx}

\title[Galactic Phylogenetics] 
{Galactic Phylogenetics}

\author[Paula Jofr\'e \& Payel Das]   
{Paula Jofr\'e$^{1,2}$
 \and Payel Das$^3$}

\affiliation{$^1$Institute of Astronomy, University of Cambridge, Madingley Road, Cambridge CB3 0HA, UK \\ email: {\tt pjofre@ast.cam.ac.uk} \\[\affilskip]
$^2$N\'ucleo de Astronom\'ia, Universidad Diego Portales, Santiago, Chile \\
$^3$Rudolf Peierls Centre for Theoretical Physics, University of Oxford, OX1 3NP, UK \\email: {\tt payel.das@physics.ox.ac.uk}}

\pubyear{2017}
\volume{334}  
\jname{Rediscovering our Galaxy}
\editors{C. Chiappini, I. Minchev, E. Starkenburg \& M. Valentini.}
\begin{document}

\maketitle

\begin{abstract}
Phylogenetics is a widely used concept in evolutionary biology. It is the reconstruction of evolutionary history by building trees that represent branching patterns and sequences. These trees represent shared history, and it is our intention for this approach to be employed in the analysis of Galactic history. In Galactic archaeology the shared environment is the interstellar medium in which stars form and provides the basis for tree-building as a methodological tool. 

Using elemental abundances of solar-type stars as a proxy for DNA, we built in \cite{Jofre17} such an evolutionary tree to study the chemical evolution of the solar neighbourhood.  In this proceeding we summarise these results and discuss future prospects. 
\keywords{Galaxy: evolution}
\end{abstract}

\firstsection 

\section{Introduction}

\subsection{Stellar DNA}
In the widely-read review of \cite{Freeman02} a very important concept was discussed: chemical tagging. Unlike the kinematical memory of long-lived low-mass stars,  the chemical pattern imprinted in their atmospheres remains intact, reflecting the chemical composition of the gas from which they formed. Hence,  the chemical abundances of stars can be used to identify the clouds from which they formed. By doing this for stars at different locations and of different ages, and complementing this information  with their kinematic properties, one can constrain chemodynamical models of the Galaxy. 

This idea, combined with the arrival of Gaia data, is motivating the development of very large high resolution spectroscopic surveys, able to provide us with about 20 elemental abundances for thousands of stars. This is further leading to the development of sophisticated clustering techniques which are able to classify different stellar populations in chemical space, and hence, identify the origins of different stellar populations. 

\subsection{Chemical continuity}
It is already well known that there is a continuity in chemical patterns between successive stellar generations. Massive stars pollute the interstellar medium with more metals enabling the formation of new stars that are more metal-rich. This implies that the origins of stars identified with chemical tagging are related to each other, and understanding their relationship is what reveals the chemical evolution of the Milky Way. 
 
\section{Evolutionary tree of solar neighborhood stars}
If we can identify the origins of stars using the concept of chemical tagging, phylogenetics offers a powerful way to complement chemical tagging and study the chemical evolution of the Milky Way. We can use the chemical pattern of stars as DNA and build evolutionary trees, in which every branch represents a different stellar population.  At a first stage, this is doing the same as chemical tagging as  we are essentially  classifying stars in their chemical space. But at a second stage, the branches can be used to study their relationships and reconstruct their shared history.  

In \cite{Jofre17} we took the sample of solar twins of \cite{Nissen16}, which comprises accurate chemical abundances of 17 different elements, ages and kinematic properties. With these abundances we constructed an evolutionary tree and found three different branches. By analysing the ages and kinematics of the stars in these branches we could attribute these branches to the thick disk, thin disk and an intermediate population. 

By further analysing their branch lengths, we could estimate a total chemical enrichment rate for each of the populations, finding that the thick disk had a faster star formation rate than the thin disk, confirming, in a purely empirical manner, previous findings. We finally identified nodes in the tree which split into multiple branches, discussing possible extreme events in the past which might drive independent evolutionary paths for different populations.  Although our sample of stars was very small, we showed that this approach has great potential to disentangle the different physical processes that formed our Galaxy.

\section{Future}

Phylogenetic tools have existed for over a century and are based on using evolutionary trees to understand the evolution of systems. These can be biological, but also sociological (languages, religions). While the mechanisms of evolution differ, the manner in which phylogenetic trees are interpreted is remarkably similar. It is our contention that we can apply theories of Evolution to the Milky Way.   As long as we believe chemical tagging can work, we can do {\it Galactic Phylogenetics} and reconstruct the history of our Galaxy.

\end{document}